\pdfoutput=1
\documentclass[11pt,a4paper]{article}
\usepackage{jheppub}
\usepackage{amsfonts,amssymb,amsmath,mathtools,mathrsfs,slashed}
\usepackage{graphicx}		% to insert figures
\usepackage{hyperref}		% PDF hyperreferences
\usepackage{array}
\usepackage[utf8]{inputenc}
\usepackage[shortlabels]{enumitem}
\usepackage{physics}
\usepackage{color}
\usepackage{comment}
\usepackage[english]{babel}
\usepackage{bm}
\usepackage{dsfont}
\usepackage{gensymb}
\usepackage{float} %to force figures with the assistance of [H]

\newcommand{\be}{\begin{equation}}
\newcommand{\ee}{\end{equation}}
\newcommand{\bea}{\begin{eqnarray}}
\newcommand{\eea}{\end{eqnarray}}

\makeatletter
\gdef\@fpheader{}
\makeatother

\begin{document}

\preprint{HIP-2024-24/TH}

\title{Dynamics of a Higgs phase transition in the Klebanov--Witten theory}

\author[a,c]{Oscar Henriksson,}
\author[b,c]{Niko Jokela,}
\author[b,c]{Julia Junttila}

\affiliation[a]{Faculty of Science and Engineering, {\AA}bo Akademi University,\\ Henrikinkatu 2, FI-20500 Turku, Finland}
\affiliation[b]{Department of Physics, University of Helsinki,\\ P.O.~Box 64, FI-00014 University of Helsinki, Finland}
\affiliation[c]{Helsinki Institute of Physics,\\ P.O.~Box 64, FI-00014 University of Helsinki, Finland}

\emailAdd{oscar.henriksson@abo.fi}
\emailAdd{niko.jokela@helsinki.fi}
\emailAdd{julia.junttila@helsinki.fi}

\abstract{We study the dynamics of a first-order phase transition in a strongly coupled gauge theory at non-zero temperature and chemical potential, computing nucleation rates and wall speeds from first principles. The gauge theory is the four-dimensional superconformal $SU(N)\times SU(N)$ Klebanov--Witten theory, which at low temperatures displays an instability to forming scalar condensates that higgses the theory. The computation is made possible by utilizing the gravity dual, type IIB string theory on asymptotically AdS$_5\times T^{1,1}$ spacetimes. The instability is detected through the nucleation and subsequent localization of D-branes in the bulk, which in the probe limit is amenable to calculations. The nucleation rates suggest a preference for greater asymmetry between the two gauge groups as the temperature is lowered beyond its critical value. The wall speed stays below the speed of sound of the conformal plasma across all parameter values and exhibits nonlinear dependence on the energy difference between the two phases.
}

\maketitle

%%%%%%%%%%%%%%%%%%%%%%%%%%%%%%%
%%%%%%%%%%%%%%%%%%%%%%%%%%%%%%%
\section{Introduction}
%%%%%%%%%%%%%%%%%%%%%%%%%%%%%%%
%%%%%%%%%%%%%%%%%%%%%%%%%%%%%%%

First-order phase transitions (FOPTs) are ubiquitous in nature and across all areas of physics. The energy barrier between the two coexisting phases around the critical temperature prevents the field, representing the order parameter, from completing the transition in a homogeneous manner; instead, the transition often takes place through local fluctuations (quantum or thermal) which form bubbles that reach across the barrier. Once nucleated, these bubbles can start to expand, eventually converting the whole system to the new phase. The expansion of the bubbles is a complicated far-from-equilibrium process which is difficult to attack theoretically. The nucleation of the bubbles also involves significant theoretical uncertainties.

One area where the detailed dynamics of FOPTs can be of great importance is early-universe cosmology \cite{Hindmarsh:2020hop}. There, the expanding bubbles would, upon colliding, source gravitational waves, which may be observable in planned gravitational wave observatories such as LISA \cite{Caprini:2019egz,Croon:2024mde}. Such an observation would be a signal of new physics, since the Standard Model of particle physics does not give rise to thermal first-order transitions \cite{Stephanov:2006zvm,Kajantie:1995kf,Csikor:1998eu}.

FOPTs are also of interest in gauge theories at non-zero density. Quantum chromodynamics (QCD) is known to transition from a nuclear matter phase to a quark matter phase as the baryon density is increased, but the order of this transition is difficult to determine. At very high density QCD is weakly coupled and computations have revealed that the dominant low-temperature phase is likely a \emph{color superconductor} \cite{Alford:2007xm}; a Higgs phase where quarks near the Fermi surface pair up and condense. The situation at intermediate densities, including properties of possible phases and phase transitions, is unclear. This is of particular interest since astrophysical observations of neutron stars can uncover aspects of this part of the QCD phase diagram, for example the equation of state and the existence of a FOPT \cite{Annala:2023cwx}.

In this paper, we study a strongly interacting toy model which incorporates several of these themes. The model, when placed at non-zero density, can describe a (metastable) Higgs phase, reached from the high-temperature normal phase through bubble nucleation.  We investigate the dynamics of this transition, focusing on two key details: the initial nucleation of the bubbles, and their late-time, steady-state expansion. In the former case, we find critical bubbles --- saddle point solutions whose action gives the leading contribution to the nucleation rate. In the latter case, we compute the terminal velocity of the bubble wall, an important but hard-to-compute parameter entering into, for example, gravitational wave predictions. 

The toy model in question is the superconformal Klebanov--Witten (KW) theory \cite{Klebanov:1998hh}, which we are able to study at strong coupling through its holographic dual, type IIB string theory on (asymptotically) $AdS_5\times T^{1,1}$. The string theory construction was arrived at by studying a stack of D3-branes in the vicinity of a conifold singularity. In \cite{Herzog:2009gd}, gravitational solutions describing this theory at non-zero density were constructed; in \cite{Henriksson:2019ifu}, these solutions were shown to be unstable to the nucleation of both D3- and D5-branes for sufficiently low temperatures. The nucleation of such branes correspond to scalar operators in the field theory side obtaining non-zero expectation values, higgsing the gauge theory. The D3-brane instability was analyzed in \cite{Henriksson:2021zei}. Here we turn to the D5-brane case, which is slightly more involved, but also more interesting. In this case, both minima of the corresponding effective potential are at finite field values, matching more closely with a typical phase transition with two well-defined phases. (In the case of the D3-brane effective potential, in contrast, the global low-$T$ minimum was at infinity, possibly leading to a runaway behavior.) 

Holographic duality has become a popular approach for studying bubble nucleation at strong coupling, motivated in large part by future gravitational wave observatories. While computations of for example the wall speed have been done with more traditional methods at weak and, in some cases, strong coupling, (\textit{e.g.}, \cite{Moore:1995ua,Bodeker:2017cim,Li:2023xto,Yuwen:2024hme}) the prevailing difficulty motivates novel approaches. The duality allows for the computation of the field theory effective action~\cite{Ares:2021ntv}, and from that, nucleation rates and related parameters. It is also possible to methodically extract wall speeds, usually through numerical relativity simulations. A sample of previous literature is \cite{Ares:2020lbt,Bigazzi:2020phm,Bigazzi:2020avc,Ares:2021ntv,Ares:2021nap,Morgante:2022zvc,Chen:2022cgj,Chen:2024pyy}; the wall speed in particular was studied in \cite{Bea:2021zsu,Bigazzi:2021ucw,Henriksson:2021zei,Bea:2022mfb,Janik:2022wsx,Evans:2024ilx}. In this paper, we build on previous results to provide a first-principles calculation of both nucleation rates and wall speeds, without heavy numerical simulations or uncontrolled approximations. The approach can be extended to other holographic theories with similar brane nucleation/Higgs transitions, which have potential as models of new physics in the early universe and in high-density nuclear matter.

Throughout this paper, we work in the probe approximation, where the single nucleating D5-brane is assumed not to backreact significantly on the background geometry. The latter is sourced by $N$ D3-branes, where, for the semi-classical gravity limit to be a good description, $N$ should be a large number. The neglect of backreaction is not a problem for our analysis; it does however mean that the resulting geometry, with only one nucleated brane, is still unstable to further nucleation. To find a stable end state, one must eventually include backreaction, at the latest when a number of order $N$ branes have nucleated. The construction of such a Higgs state has been achieved in a bottom-up model in \cite{Henriksson:2022mgq}, and in the case of $\mathcal{N}=4$ super-Yang--Mills theory in \cite{Kim:2023sig,Choi:2024xnv}. 

The rest of the paper is organized as follows: In section \ref{sec:fieldTheoryGravityDual}, we introduce the KW theory and its gravitational dual in more detail, and describe the specific charged black brane geometries we study. In section \ref{sec:D5embedding}, we embed D5-branes, including dissolved D3-branes, in these geometries, and derive an effective action describing their dynamics. In section \ref{sec:nucleation}, we use the effective action to find critical bubble solutions, allowing us to extract the leading contribution to the bubble nucleation rate. In section \ref{sec:wallSpeed}, we compute the speed of the brane bubble wall, as well as its profile, at late times. Finally, we conclude and discuss future directions in section \ref{sec:discussion}.

%%%%%%%%%%%%%%%%%%%%%%%%%%%%%%%
%%%%%%%%%%%%%%%%%%%%%%%%%%%%%%%
\section{Field-theoretical model and its gravity dual}\label{sec:fieldTheoryGravityDual}
%%%%%%%%%%%%%%%%%%%%%%%%%%%%%%%
%%%%%%%%%%%%%%%%%%%%%%%%%%%%%%%

We will study phase transitions in the KW theory, a superconformal four-dimensional field theory with a well-understood string theory dual. The KW theory has $\mathcal{N}=1$ supersymmetry, and an $SU(N) \times SU(N)$ gauge group. The matter content consists of chiral multiplets $A_\alpha$ and $B_{\tilde\alpha}$, $\alpha, \tilde\alpha=1,2$, which transform the bifundamental $(N,\bar N)$ and $(\bar N,N)$ representations, respectively. The theory has the global symmetry $SU(2)_A\times SU(2)_B\times U(1)_R\times U(1)_B$; the $A$- and $B$-fields are each doublets under their own $SU(2)$, being singlets under the other, and carry R-charge 1/2. We will be particularly interested in the last $U(1)$-factor, with subscript $b$ for baryon, under which the $A$- and $B$-fields carry \emph{opposite} charge. The baryon label comes from the fact that gauge invariant operators of dimension of order $N$ --- baryonic operators, such as $\det A_\alpha$ --- are charged under it, while operators of order 1 --- mesonic operators, such as $\text{Tr} A_\alpha B_{\tilde\alpha}$ --- are uncharged. 

The KW theory is realized in type IIB string theory by a stack of $N$ D3-branes placed at the tip of the so-called conifold, a cone with base manifold $T^{1,1}$, with isometry group $SU(2)_A\times SU(2)_B\times U(1)_R$. This almost matches the global symmetries of the field theory; the remaining baryonic symmetry originates from the dimensional reduction of the four-form Ramond-Ramond gauge potential $C_4$. The five-form flux $F_5=\dd C_4$ is quantized, requiring
\begin{equation}
    \frac{L^4}{\alpha'^2}=\frac{27\pi}{4} g_s N \ ,
\end{equation}
where $\alpha'$ is the Regge slope, $L$ is the curvature radius, and $g_s$ is the string coupling.
As is common, we will focus on the large-$N$, strong coupling limit, where the dual string theory reduces to classical supergravity.

We consider this theory at non-zero temperature $T$, and with a non-zero chemical potential $\mu$ for the baryon charge. Since the theory is conformal, the physics only depends on the dimensionless ratio $T/\mu$. At high $T/\mu$, a normal disordered plasma phase dominates. However, as the temperature is lowered, the normal phase becomes unstable to a type of finite-density Higgs mechanism, as we will see in detail in section \ref{sec:D5embedding}.

%%%%%%%%%%%%%%%%%%%%%%%%%%%%%%%
%%%%%%%%%%%%%%%%%%%%%%%%%%%%%%%
\subsection{Background geometry}
%%%%%%%%%%%%%%%%%%%%%%%%%%%%%%%
%%%%%%%%%%%%%%%%%%%%%%%%%%%%%%%

Let us now review the classical gravity dual of the normal plasma phase of KW theory at large-$N$, strong coupling, and non-zero $T$ and $\mu$. Following \cite{Herzog:2009gd}, we take the five-form field strength to be $F_5=L^4(\mathcal{F}+\ast \mathcal{F})$, where $\ast$ is the 10D Hodge operation, with
\begin{equation}\label{eq:F5}
    \mathcal{F} = -\frac{2}{27}\omega_2\wedge\omega_2\wedge g_5 - \frac{1}{9\sqrt{2}}\dd A\wedge\omega_2\wedge g_5 \ .
\end{equation}
Here, we have used the following forms defined on the internal space:
\begin{equation}
\begin{split}
\omega_2 &\equiv \frac{1}{2} \left(\sin \theta_1 \dd\theta_1 \wedge \dd\phi_1
- \sin \theta_2 \dd\theta_2 \wedge \dd\phi_2 \right)  \\
g_5 &\equiv \dd \psi + \cos \theta_1 \dd\phi_1 + \cos \theta_2 \dd\phi_2 \ .
\end{split}
\end{equation}
In (\ref{eq:F5}), the first term gives the standard $N$ units of flux through $T^{1,1}$. The second term introduces the one-form $A$ which, upon dimensional reduction, becomes the $U(1)$ gauge field dual to the baryon current.

This extra gauge field causes the metric to deviate from pure AdS$_5\times T^{1,1}$ through the squashing of the internal space. This is captured by a 10D metric of the form
\begin{equation}\label{eq:10Dmetric}
 ds_{10}^2 = L^2 e^{-5 \chi/3} \dd s_5^2 + L^2e^{\chi}
\bigg[{e^{\eta} \over 6} (\dd \theta_1^2 + \sin^2 \theta_1 \dd\phi_1^2)
+ {e^{\eta} \over 6} (\dd \theta_2^2 + \sin^2 \theta_2 \dd\phi_2^2)
+ {e^{-4 \eta} \over 9} g_5^2 \bigg] \ .
\end{equation}
Here, $\dd s_5^2$ is a 5D asymptotically-AdS line element, $(\theta_i,\phi_i)$ with $i=1,2$ are two sets conventional spherical coordinates, and $\psi\in [0,4\pi)$. We note that the internal metric, within the square brackets, takes the form of a $U(1)$ fibration over $S^2\times S^2$. The scalars $\eta$ and $\chi$ realize the aforementioned squashing.

Carrying out the dimensional reduction from 10D to 5D yields a theory with Lagrangian
\begin{equation}
    \mathcal{L}_{5D} = R - \frac{10}{3}(\partial_\mu \chi)^2 - 5(\partial_\mu \eta)^2 - V(\chi,\eta) - \frac{1}{4}e^{2\eta-\frac{4}{3}\chi}F_{\mu\nu}^2 \ ,
\end{equation}
with indices $\mu,\nu=0,\ldots,4$, field strength $F=\dd A$ (with $A$ the above gauge field), and scalar potential
\begin{equation}
    V(\chi,\eta) = \frac{8}{L^2}e^{-\frac{20}{3}\chi} + \frac{4}{L^2}e^{-\frac{8}{3}\chi}\left( e^{-6\eta} - 6e^{-\eta} \right) \ .
\end{equation}
The black-brane solutions we are interested in can be described by the following {\emph{Ansatz}} for the 5D metric:
\begin{equation}\label{eq:5Dmetric}
\dd s_5^2 = - g(r) e^{-w(r)} \dd t^2 + \frac {\dd r^2}{g(r)} + r^2 \dd\vec x_3^2 \ .
\end{equation}
Here, $g(r_H)=0$ at the horizon radius $r_H$. Assuming also that the scalar fields only have radial dependence, and that $A=\Phi(r)dt$, the resulting equations of motion read
\begin{align}
    \chi''+\chi'\left( \frac{3}{r}+\frac{g'}{g}+\frac{5r}{3}\eta'+\frac{10r}{9}\chi'^2 \right) -\frac{\Phi'^2}{10g}e^{w+2\eta-\frac{4}{3}\chi} - \frac{3}{20g}\frac{\partial V}{\partial\chi} =&\ 0  \\
    \eta''+\eta'\left( \frac{3}{r}+\frac{g'}{g}+\frac{5r}{3}\eta'+\frac{10r}{9}\chi'^2 \right) +\frac{\Phi'^2}{10g}e^{w+2\eta-\frac{4}{3}\chi} - \frac{1}{10g}\frac{\partial V}{\partial\eta} =&\ 0  \\
    g'+g\left( \frac{2}{r}+\frac{5r}{3}\eta'^2+\frac{10r}{9}\chi'^2 \right) +\frac{r}{6}\Phi'^2e^{w+2\eta-\frac{4}{3}\chi}+\frac{r}{3}V =&\ 0  \\
    \Phi'' +\Phi'\left( \frac{3}{r}+\frac{w'}{r}+2\eta'-\frac{4}{3}\chi' \right) =&\ 0  \\
    w'+\frac{10r}{9}\left( 3\eta'^2+2\chi'^2 \right) =&\ 0 \ .
\end{align}
It is straightforward to see that these admit a pure AdS$_5$ solution, with $\chi=\eta=0$, $\Phi=w=\text{constant}$, and $g=(r/L)^2$, which will furnish the large-radius UV-asymptotics of our geometries.\footnote{We choose our time units such that $w\to 0$ as $r\to\infty$.} Expanding our equations around this solution at $r\to\infty$ shows that the scalar fields $\chi$ and $\eta$ fall off as $r^{-8}$ and $r^{-6}$, respectively, and are thus dual to irrelevant operators. The gauge field $\Phi$, on the other hand, falls of as
\begin{equation}
    \Phi(r) = \mu + \frac{Q}{r^2} + \ldots \ , \ \ r\to\infty \ ,
\end{equation}
where $\mu$ is determines the chemical potential, and $Q$ the charge density, of the dual baryon current.

The full asymptotically-AdS charged black brane solutions were constructed numerically (using a shooting method) in \cite{Herzog:2009gd}, where the interested reader can find more details. The states dual to these solutions have a temperature $T$ and an entropy density $s$ determined by the gravity solution evaluated at the black hole horizon:
\begin{equation}
    T=\frac{1}{4\pi}e^{-\frac{w(r_H)}{2}}g'(r_H) \qquad \text{and} \qquad s=\frac{r_H^3}{4G_5}
\end{equation}
Here, $r_H$ denotes the horizon radius, and $G_5$ is the 5D gravitational constant. In the rest of the paper, we rescale the radial coordinate such that the horizon is always at $r_H=1$. The solutions/states form a one-parameter family, which can be parameterized by the dimensionless ratio $T/\mu$. Here, we are interested in the behavior of D5-branes probing this geometry, to which we now turn.

%%%%%%%%%%%%%%%%%%%%%%%%%%%%%%%
%%%%%%%%%%%%%%%%%%%%%%%%%%%%%%%
\section{D5-brane embedding and effective action}\label{sec:D5embedding}
%%%%%%%%%%%%%%%%%%%%%%%%%%%%%%%
%%%%%%%%%%%%%%%%%%%%%%%%%%%%%%%

It is clearly interesting to add D5-branes to the D3-brane conifold system. By wrapping two of the dimensions spanned by D5-branes around an $S^2$ in the internal manifold ($T^{1,1}$ is topologically $S^2\times S^3$) the remaining 3+1 dimensions act as a domain wall in the external part of the spacetime. In \cite{Gubser:1998fp}, it was argued that this domain wall enforces a jump of unity in the rank of \emph{one} of the two gauge groups: $SU(N)\times SU(N)\to SU(N)\times SU(N+1)$, while a D3-brane domain wall would lead to changing the rank of \emph{both} groups.  Bringing such a wrapped D5-brane down to the singular point of the conifold, where the two-sphere collapses, it is termed a \emph{fractional} D3-\emph{brane}. Such a construction leads to the famous Klebanov--Strassler theory \cite{Klebanov:2000hb}, which exhibits many interesting phenomena including a duality cascade and chiral symmetry-breaking.

In this paper, we want to keep the conformal KW theory as our UV geometry, so we cannot simply add D5-branes haphazardly. However, one can imagine adding a D5- \emph{and} an anti-D5-brane, such that the total fluxes at the AdS-boundary remain unchanged. In a non-trivial background geometry, the two branes might be pulled apart, one of them moving towards the AdS boundary and the other towards the black brane, vanishing from sight at the horizon. If energetically favorable, this process can take place spontaneously, much like a higher-dimensional version of Schwinger pair production of charged particles in an electric field. If a stable end state is reached, it should represent a Higgs phase of the dual field theory. 

In \cite{Henriksson:2019ifu}, it was shown that such instabilities are indeed present in the charged plasma described by the black brane solutions of the previous section. The instability is present for both D3- and D5-branes, at $T/\mu\lesssim 0.2$. One can compare this with similar instabilities found in (the holographic dual of) $\mathcal{N}=4$ super-Yang-Mills at finite R-charge \cite{Yamada:2008em,Henriksson:2019zph}. There, when the field theory is defined on $\mathbb{R}^3$, the instability occurs for any finite chemical potential. In that case, the branes in question are charged under the R-symmetry (they rotate in the internal space), and the instability can be thought of as a consequence of centrifugal forces on the branes. In the present case of KW theory, on the other hand, the branes we study carry no baryon charge, and the reason for the instability is less direct. In \cite{Henriksson:2021zei}, the KW theory instability for D3-branes was studied in some detail. Here, we turn to the D5-case.

To embed the probe D5-brane, we parametrize the worldvolume of the brane by the spacetime coordinates $\{t,x,y,z,\theta_1,\phi_1\}$. Following \cite{Dasgupta:1999wx}, we take the brane to wrap the $S^2$ of the conifold which corresponds to ``$S^2_1-S^2_2$'', where $S^2_1$ and $S^2_2$ are the two-spheres parameterized by $(\theta_1,\phi_1)$ and $(\theta_2,\phi_2)$, respectively. Thus, we let $\Theta_2=\theta_1$ and $\Phi_2=-\phi_1$. The brane will be located on a constant, and arbitrary, $\psi$-coordinate. In principle, the brane could be allowed to move in the $\psi$-direction; however, such motion only increases the energy, and we are interested in configurations which \emph{lower} the total energy. Lastly, we allow a very general embedding $R=R(t,\vec x)$ in the radial direction. In the following, indices $i,j,\ldots$ run over the three spatial field theory coordinates, and we denote $t$-derivatives by a dot. The 10D metric will be denoted by $G$, with components $G_{\mu\nu}$.

The probe D5-brane action in our background will contain the standard DBI term, plus a Wess-Zumino (WZ) term coupling to potential $C_4$ of the five-form field ($F_5=\dd C_4$) through the introduction of a worldvolume flux $\mathcal{F}$:
\begin{equation}\label{eq:D5action}
  S_{D5} = S_{DBI} + S_{WZ} = -T_5\int \dd^6\xi \sqrt{-\det (P[G]+\mathcal{F})} + T_5\int P[C_4]\wedge \mathcal{F}
\end{equation}
Here $P[\dots]$ denotes the pullback of spacetime fields to the brane worldvolume. The constant dilaton has been absorbed into the tension of the brane, given by \cite{Herzog:2009gd}
\begin{equation}\label{eq:tension}
 T_5 = \frac{1}{(2\pi)^5 g_s l_s^6} = \frac{81\sqrt{3}}{256\pi^{7/2}}\frac{\sqrt{g_s}N^{3/2}}{L^6} = \frac{81\sqrt{3}}{256\pi^{7/2}}\frac{\sqrt{\lambda}\, N}{L^6} \ ,
\end{equation}
where in the last step we introduced the 't Hooft-coupling $\lambda=g_s N$. The worldvolume flux $\mathcal{F}$ will be taken to be
\begin{equation}
    \mathcal{F} = L^2 q \sin{\theta_1} \dd\theta_1 \wedge \dd\phi_1 \ .
\end{equation}
Such a flux couples to $C_4$ and thus acts as a D3-brane charge, effectively \emph{dissolving} a number of D3-branes in the D5-brane \cite{Douglas:1995bn}. In fact, this flux is quantized,
\begin{equation}\label{eq:fluxQuantization}
    q=\frac{\pi\alpha' M}{L^2} = \frac{2\sqrt{\pi}}{3\sqrt{3}}\frac{M}{\sqrt{\lambda}} \ ,
\end{equation}
where the integer $M$ gives the number of dissolved D3-branes \cite{Arean:2006vg}. Later we will be considering $q$-values of order one, meaning a number of dissolved D3-branes of order $\sqrt{\lambda}$. Since we are working in the limit of large-$\lambda$, $q$ can be treated as a continuous variable.

We now turn to the induced metric on the brane, which can be written as
\begin{equation}
 \dd s^2_4 = \gamma_{tt} \dd t^2 + 2\gamma_{ti} \dd t\, \dd x^i + \gamma_{ij} \dd x^i \dd x^j \ ,
\end{equation}
where the components are given by
\begin{equation}
\begin{split}
 \gamma_{tt} &= G_{tt} + G_{rr} \dot R^2 \\
 \gamma_{ti} &= G_{rr}\dot R\, (\partial_i R) \\
 \gamma_{ij} &= G_{ij} + G_{rr}(\partial_i R)(\partial_j R) \ .
\end{split}
\end{equation}
Computing the DBI-part of the action, we find
\begin{multline}
 S_{DBI} = -L^6 T_5 \int \dd t\, \dd\vec x\, \dd\theta_1 \dd\phi_1 \left(\rule{0cm}{9mm} R^3 e^{-\frac{w}{2}-\frac{10}{3}\chi}\sin{\theta_1} \times \right. \\ 
 \left. \times\sqrt{ \left( g-\frac{e^w}{g}\dot R^2 + \frac{1}{R^2} \sum_i(\partial_i R)^2 \right) \left( \frac{e^{2\chi+2\eta}}{9} + q^2 \right) } \, \rule{0cm}{9mm}\right)
\end{multline}
Note that here the scalar and metric functions ($\chi,\eta,w,g$) are functions of $R(t,\vec x)$.

For the WZ term of the action, we need the pullback of $C_4$. The relevant component, which couples to the D5-branes we study here, is $L^4 a_4(r)\, \dd t\wedge \dd x_1\wedge \dd x_2\wedge \dd x_3$, where $a_4(r)$ is found by integrating $a_4'(r) = 4r^3e^{-\frac{w}{2}-\frac{20}{3}\chi}$ with the condition that it goes to zero on the horizon. Thus we can write it as
\begin{equation}
 a_4(r) = 4\int_{r_H}^r \dd \tilde{r} \left(\tilde{r}^3e^{-\frac{w(\tilde{r})}{2}-\frac{20\chi(\tilde{r})}{3}}\right) \ .
\end{equation}
The WZ-part of the action is then
\begin{equation}
 S_{WZ} = L^6 T_5 \int \dd t\, \dd\vec x\, \dd\theta_1 \dd\phi_1 \left\{ a_4(R)\, q \sin{\theta_1} \right\} \ .
\end{equation}
We can carry out the trivial integral over the angular directions, and use  (\ref{eq:tension}), to arrive at the following effective D5-brane action,
\begin{equation}\label{eq:D5actionComplete}
    S_{D5} = 4\pi L^6 T_5 \int \dd t\, \dd\vec x \, \mathcal{L}_{D5}  = \frac{81\sqrt{3}}{64\pi^{5/2}}\sqrt{\lambda}\, N \int \dd t\, \dd\vec x \, \mathcal{L}_{D5} \ ,
\end{equation}
where we have defined the effective Lagrangian density
\begin{equation}\label{eq:D5effLagrangian}
\mathcal{L}_{D5} \equiv -\left\{ R^3 e^{-\frac{w}{2}-\frac{10}{3}\chi}\sqrt{ \left( g-\frac{e^w}{g}\dot R^2 + \frac{1}{R^2} \sum_i(\partial_i R)^2 \right) \left( \frac{e^{2\chi+2\eta}}{9} + q^2 \right) } - q\, a_4(R) \right\} \, .
\end{equation}
As noted in \cite{Henriksson:2021zei} in the case of D3-branes in these conifold backgrounds, such an action can be interpreted as a quantum effective action for a scalar operator in the dual KW theory. If a D5-brane would localize at some specific radius in the bulk, the dual operator acquires an expectation value, thereby higgsing the gauge theory.

In the next section, we will compare the nucleation rate of D5-branes with those of D3-branes computed in \cite{Henriksson:2021zei}. For now, we just note that in the limit of large flux $q$, the D5 is dominated by the presence of the dissolved D3-branes, and $S_{D5}$ becomes
\begin{equation}\label{eq:D5actionD3limit}
 S_{D5}\to -\frac{81\sqrt{3}}{64\pi^{5/2}}\sqrt{\lambda}\, N\, q \int \dd t\, \dd\vec x \Bigg\{ R^3 e^{-\frac{w}{2}-\frac{10}{3}\chi}\sqrt{ \left( g-\frac{e^w}{g}\dot R^2 + \frac{1}{R^2} \sum_i(\partial_i R)^2 \right) } - a_4(R) \Bigg\} \ .
\end{equation}
Comparing with the D3-brane action in the same geometry (see \cite{Henriksson:2021zei}), we note that this corresponds to the action of $M$ D3-branes as long as the previously mentioned quantization condition (\ref{eq:fluxQuantization}) holds.

\begin{figure}[H]
    \centering
    \includegraphics[scale = 0.72]{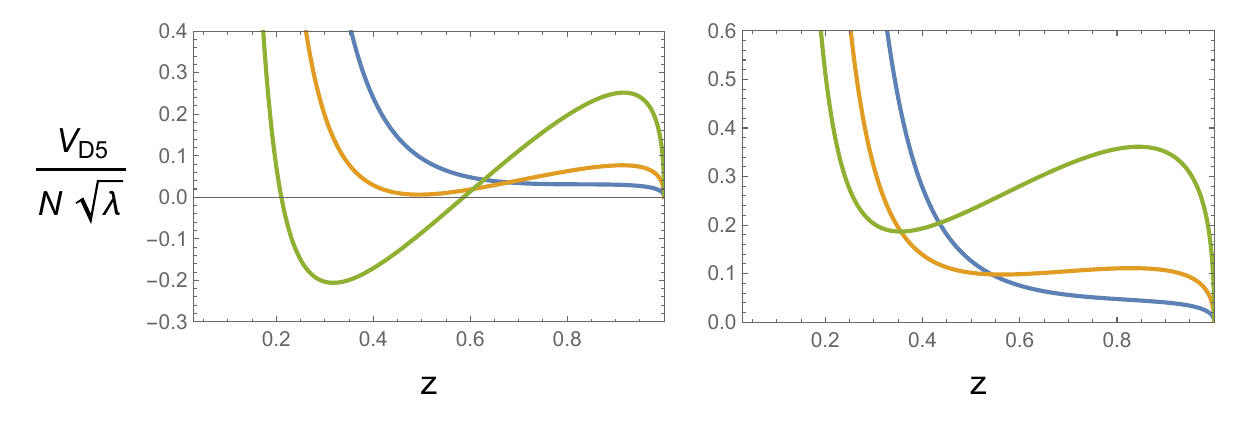}
\caption{The D5-brane effective potential as a function of the radial coordinate $z=r_H/r$. On the left, $T/\mu\approx 0.12$ which is in the unstable region. On the right, $T/\mu\approx 0.21$ which is in the stable region. The blue, orange, and green curves correspond to $q=1,3,10$, respectively.}\label{fig:D5potentialVSz}
\end{figure}

The effective potential for a D5-brane is obtained by setting all derivatives of $R$ in the action to zero, giving
\begin{equation}\label{eq:D5effPot}
 V_{D5} = \frac{81\sqrt{3}}{64\pi^{5/2}}\sqrt{\lambda}\, N \left( R^3 e^{-\frac{w}{2}-\frac{10}{3}\chi}\sqrt{ g \left( \frac{e^{2\chi+2\eta}}{9} + q^2 \right) } - q\, a_4(R) \right) \ .
\end{equation}
As discussed in \cite{Henriksson:2019ifu}, for $T/\mu\lesssim 0.2$ one finds a critical value of the flux, $q=q_c(T/\mu)$, above which the effective potential has a global minimum outside the horizon at some radius $r=r_\text{min}$, indicating an instability to D5-brane nucleation. In Fig.~\ref{fig:D5potentialVSz}, the potential is plotted for $T/\mu\approx 0.12$ (unstable) and $T/\mu\approx 0.21$ (stable), as a function of the radial coordinate $z=r_H/r$. In the former case, we see that the orange curve, with $q=3$, is at the edge of stability; larger flux, such as the green curve with $q=10$, give an instability. In the stable case, on the other hand, the potential never dips below zero for any value of the flux.

%%%%%%%%%%%%%%%%%%%%%%%%%%%%%%%
%%%%%%%%%%%%%%%%%%%%%%%%%%%%%%%
\section{Brane bubble nucleation}\label{sec:nucleation}
%%%%%%%%%%%%%%%%%%%%%%%%%%%%%%%
%%%%%%%%%%%%%%%%%%%%%%%%%%%%%%%

We now use the above effective action to find critical bubbles, static solutions to the EoM that asymptote towards the ``false vacuum" at the horizon, and whose action give the leading contribution to the D5-brane nucleation rate. We will study the dependance of the action on the parameters $T/\mu$ and the worldvolume flux $q$.

Critical bubbles typically have spherical symmetry, so we switch to spherical coordinates $\{\rho,\alpha,\beta\}$ on the D5-brane worldvolume and assume $R=R(\rho)$. We can then perform the angular integrals over $\alpha$ and $\beta$, giving a factor of $4\pi$, and the integral over Euclidean time, giving a factor of $1/T$:
\begin{equation}\label{eq:D5bubbleAction}
 S_{D5} = \frac{81\sqrt{3}}{16\pi^{3/2}} \frac{\sqrt{\lambda}\, N}{T} \int \dd\rho\, \rho^2 \Bigg\{ R^3 e^{-\frac{w}{2}-\frac{10}{3}\chi}\sqrt{ \left( g + \frac{(R')^2}{r^2} \right) \left( \frac{e^{2\chi+2\eta}}{9} + q^2 \right) } - q\, a_4(R) \Bigg\} \ .
\end{equation}
The EoM that follows from varying this action is a second order ordinary differential equation, which describes how the brane bubble stretches from the black brane horizon to the vicinity of the true vacuum. We solve this complicated equation numerically. The equation has two singular points which complicate our task: at center of the bubble, at $\rho=0$, and the edge of the bubble, where it touches the horizon, at some $\rho=\rho^*$ such that $R(\rho^*)=r_H=1$. We deal with this as follows: First, expand the solution near the center of the bubble, at $\rho=0$, as
\begin{equation}
    R(\rho) = R(0) + \frac{1}{2} R''(0) \rho^2 + \ldots \ .
\end{equation}
Here we have used the boundary condition that $R'(0)=0$. By plugging this expansion into the EoM, we can fix $R''(0)$, as well as all higher derivatives, in terms of $R(0)$ --- in practice, we stop at seventh order in the expansion. We then use this expansion to set boundary conditions at some small non-zero $\rho$-value, typically $10^{-3}$, thereby avoiding one of the singular points. 

To deal with the other singular point, we write a series solution valid near the horizon, of the form
\begin{equation}
 R(\rho) = 1 + R_1(\rho^*-\rho) + R_2(\rho^*-\rho)^2 + \ldots \ .
\end{equation}
Here, the EoM fixes the constants $(R_1,R_2,\ldots)$ in terms of  $\rho^*$. We then use \textit{Mathematica} and its built-in function \textbf{NDSolve} to set up a shooting method, guessing a value of $R(0)$ and integrating the EoM from $\rho=10^{-3}$ to a point near the horizon, say where $R(\rho)=1+10^{-2}$, where we attempt to match the solution and its first derivative to the near-horizon expansion. This matching requires specific values for $\rho^*$ and $R(0)$. We thereby arrive at a smooth solution valid from the center of the bubble to its edge.

\begin{figure}[H]
    \centering
    \includegraphics[scale = 0.7]{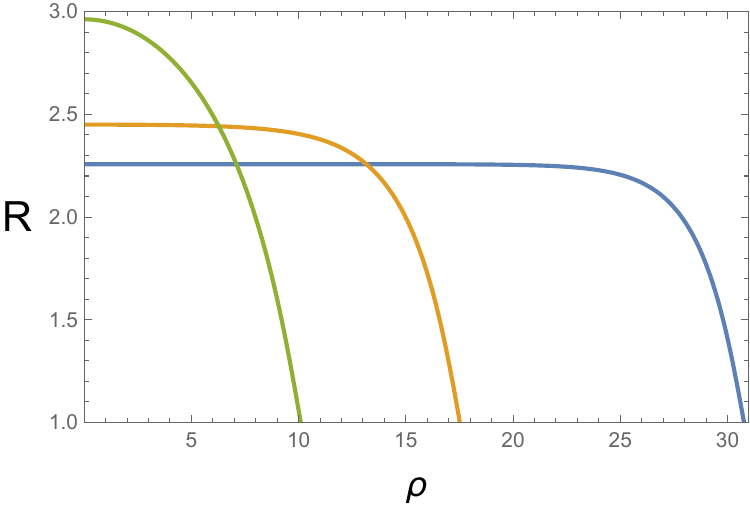}
    \caption{Example bubble solutions; the brane position in the holographic dimension as a function of ``field theory radius" $\rho$. Here, $T/\mu\approx 0.12$, and the blue, orange, and green curves are branes with flux $q=4,5$, and 20, respectively. Recall that the black brane horizon is at $R=r=1$.}
    \label{fig:bubbleSolutions}
\end{figure}

Some typical bubble solutions resulting from this procedure are shown in Fig.~\ref{fig:bubbleSolutions}, for a background with $T/\mu\approx 0.12$. As mentioned earlier, a single D5-brane cannot nucleate on its own, since it would change the asymptotic fluxes. Instead, this nucleation should be thought of as occuring together with the nucleation of a compensating anti-D5-brane, which is located exactly on the horizon at $r=1$, extending from $\rho=0$ to $\rho=\rho^*$ at the edge of the bubble. Such an anti-brane is also static, being frozen on the horizon in the boundary time $t$ that is relevant here, and contributes nothing to the bubble action.

\begin{figure}[H]
    \centering
    \includegraphics[scale = 0.74]{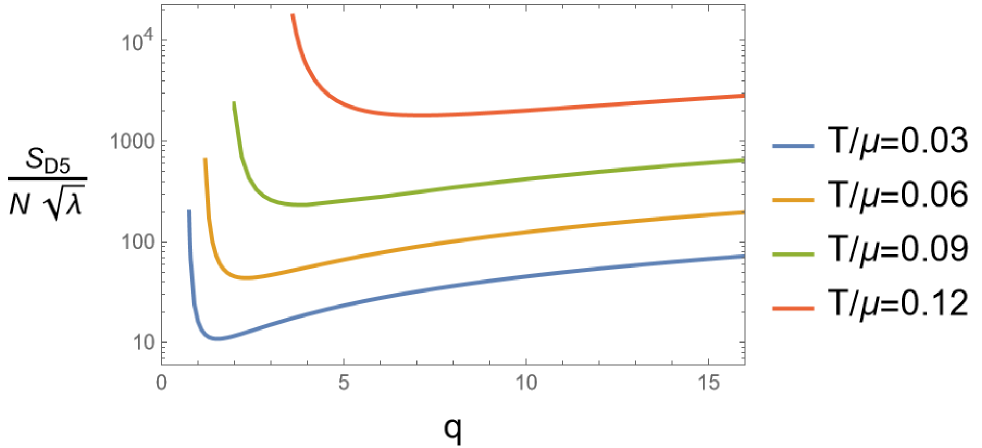}
    \caption{Bubble action as a function of the flux $q$ for different $T/\mu$.}
    \label{fig:bubbleActionVSq}
\end{figure}

\begin{figure}[H]
    \centering
    \includegraphics[scale = 0.74]{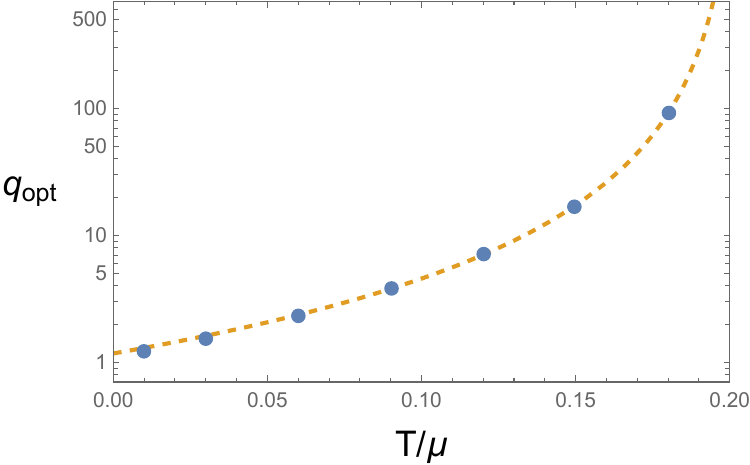}
    \caption{Blue dots represent optimal $q$-values at different $T/\mu$. The orange-dashed line is the fit to the function (\ref{eq:qOpt}).}
    \label{fig:qOptVSTmu}
\end{figure}

To compute the action, we simply plug our D5-brane solutions back into Eq. (\ref{eq:D5bubbleAction}) and integrate (numerically) over $\rho$. Some results are shown in Fig. \ref{fig:bubbleActionVSq}. For a fixed $T/\mu$, we observe that there is always a finite $q$-value that minimizes the action, and thus leads to the largest nucleation rate. This \emph{optimal} $q$-value $q_\text{opt}$ is plotted for different $T/\mu$ in Fig. \ref{fig:qOptVSTmu}.  We note that $q_\text{opt}$ diverges as we approach the critical temperature $T_c\approx 0.2\mu$, and is well fitted by the function
\begin{equation}\label{eq:qOpt}
    q_\text{opt}=\frac{a\, \mu^2}{(T-T_c)^2} \ ,
\end{equation}
with $a\approx 0.05$, as shown by the orange-dashed line in the same figure. The optimal $q$ is always close to the minimal $q$-value needed for the nucleation to take place, which also corresponds to a thin-wall limit for the bubbles. While we have not attempted it, it might be possible to derive (\ref{eq:qOpt}) analytically in this limit.

From Fig. \ref{fig:qOptVSTmu} we can also conclude that the action appears to stay finite even as the temperature approaches zero. Thus, the high-temperature plasma phase remains metastable throughout, with no sign of a spinodal point.

%%%%%%%%%%%%%%%%%%%%%%%%%%%%%%%
%%%%%%%%%%%%%%%%%%%%%%%%%%%%%%%
\subsection{Comparing with results for D3-branes}
%%%%%%%%%%%%%%%%%%%%%%%%%%%%%%%
%%%%%%%%%%%%%%%%%%%%%%%%%%%%%%%

The most important difference between the action of a D3- and a D5-brane is the factor of $\sqrt{\lambda}=\sqrt{g_s N}$. Since, as usual, we work in the holographic large-$\lambda$ limit, this means that the D5-brane action will in general be much larger than the D3-brane action, and D5-brane nucleation will be greatly suppressed compared with D3-brane nucleation.

It is interesting to ask if this might change if we take $\lambda$ finite and move towards the weak-coupling limit of the field theory (still at large $N$). It is clear that the action (\ref{eq:D5bubbleAction}) will become smaller than the D3-brane action for $\lambda$ sufficiently small; the question is if this happens at a point where our results, which assumed large $\lambda$, might still be trustworthy.

To check this, we take the value of the D5-brane action at $q=q_\text{opt}$, at each $T/\mu$. We then compare it to the value of the D3-brane action at the same $T/\mu$. Treating $\lambda$ in the D5-brane action as a free parameter, we check what value it must take for the two actions to be equal. The answer is that only for $\lambda<1$, clearly outside of the validity of holographic computations, does the D5-brane action start to dominate over the D3-brane action. At low temperature, the critical $\lambda$-value becomes larger, but still only around 0.1. Hence, D3-brane nucleation dominates everywhere in the range of validity of the large-$\lambda$-approximation. It would nonetheless be interesting to carry out the corresponding computation at weak coupling directly in the KW theory; bravely extrapolating our results, one can then hypothesize that the D5-brane instability will dominate.

%%%%%%%%%%%%%%%%%%%%%%%%%%%%%%%
%%%%%%%%%%%%%%%%%%%%%%%%%%%%%%%
\section{Terminal wall speed}\label{sec:wallSpeed}
%%%%%%%%%%%%%%%%%%%%%%%%%%%%%%%
%%%%%%%%%%%%%%%%%%%%%%%%%%%%%%%

We now move on to consider the late-time behavior of the nucleating D5-brane bubbles. The critical bubbles we found in the previous section are static but unstable solutions, which, upon a small perturbation, will either start to contract or expand. In the latter case, the accelerating bubble will asymptote towards a terminal velocity determined by the pressure difference across the bubble wall and the friction between the wall and the surrounding medium. We will now determine the terminal velocity, together with the profile of the bubble wall at late times.

After the bubble has expanded for a long time, any particular patch of it will be effectively planar, moving in (say) the $x_1$-direction. Then it is useful to parameterize the worldvolume by its $r$-coordinate instead of its $x_1$-coordinate, and work with the embedding function $X_1 \equiv X(t, r)$. This change of variables recasts the action in (\ref{eq:D5actionComplete}) as
\bea
 S_{D5} & = & -4\pi L^6 T_5 \int \dd t\, \dd r\, \dd y\, \dd z \Bigg\{ r^2 e^{-\frac{w}{2}-\frac{10}{3}\chi}\sqrt{\left( 1 -\frac{e^w r^2}{g}\dot X^2 + r^2 g (X')^2 \right) \left( \frac{e^{2\chi+2\eta}}{9} + q^2 \right) } \nonumber\\
  && \qquad\qquad\qquad\qquad\qquad\qquad - X' q\, a_4(r) \Bigg\} \ .
\eea
Note that a prime now denotes a derivative with respect to $r$. Since now only derivatives of $X$ show up in the action, one can define an associated conserved momentum current $P^\mu$, whose non-zero components are
\begin{align}
 P^t &\equiv \frac{-1}{4\pi L^6 T_5}\frac{\delta S_{D5}}{\delta \dot X} = -\frac{r^4}{3} e^{\frac{w}{2}-\frac{10}{3}\chi} \sqrt{\frac{e^{2\chi+2\eta} + 9q^2}{ g \left(g-e^w r^2\dot X^2 + r^2 g^2 X'^2\right) }}\, \dot X \\
P^r &\equiv \frac{-1}{4\pi L^6 T_5}\frac{\delta S_{D5}}{\delta X'} = \frac{r^4}{3} g\, e^{-\frac{w}{2}-\frac{10}{3}\chi} \sqrt{\frac{e^{2\chi+2\eta} + 9q^2}{ 1-\frac{e^w r^2}{g}\dot X^2 + r^2 g X'^2 }}\, X' - q\, a_4(r) \ . \label{eq:PrDefinition}
\end{align}
Note the overall constant we have factored out in the above definitions; this is convenient since it simplifies the coming equations, and we will in either case not be interested in evaluating the magnitude of the current. The conservation of this current implies
\begin{equation}\label{eq:momentumCurrentConserved}
 \partial_t P^t + \partial_r P^r = 0 \ .
\end{equation}
We now search for steady-state solutions to this equation. We do this by a series of steps similar to those in \cite{Henriksson:2021zei}, which in turn drew inspiration from the drag force computations in \cite{Gubser:2006bz, Herzog:2006gh}, called the Karch--O'Bannon method in the D-brane context~\cite{Karch:2007pd}. We begin with the {\emph{Ansatz}}
\begin{equation}
 X(t,r)=vt + \xi(r) \ ,
\end{equation}
where $v$ is the terminal velocity of the brane bubble wall. Plugging into the conserved current we find that $P^t$ is independent of $t$ and thus drops out of the equation of motion. For $P^r$ we get
\begin{equation}
 P^r = \frac{r^4}{3} g\, e^{-\frac{w}{2}-\frac{10}{3}\chi} \sqrt{\frac{e^{2\chi+2\eta} + 9q^2}{ 1-\frac{e^w r^2}{g} v^2 + r^2 g\, \xi'^2 }}\, \xi' - q\, a_4(r) \ ,
\end{equation}
which must be a constant by (\ref{eq:momentumCurrentConserved}). Solving for $\xi'(r)$ we get
\begin{align}\label{eq:D5_KsiiDeriv}
    \xi'(r) = \pm \frac{P^r+q\, a_4}{g} \sqrt{ \frac{e^{w}v^2-g/r^2}{(P^r+qa_4)^2-\frac{1}{9}r^6g\, e^{-w-\frac{20}{3}\chi}\left( e^{2\chi+2\eta}+9q^2 \right)} } \ ,
\end{align}
where we recall that $a_4, g, w, \chi $, and $ \eta$ are functions of $r$.

We will now explain how to analyze this equation in order to determine $P^r$ and, more importantly, the wall speed $v$. This is done in several steps, by requiring that the profile function $\xi(r)$ is everywhere real and has the correct shape.

First, we know that at late times, the bubble front should approach the true vacuum (global minimum) of the effective potential (\ref{eq:D5effPot}). Calling the radius where this vacuum is found $r_\text{min}$, we must then require that $\xi'(r)$ diverges at $r=r_\text{min}$. The only way this can happen is if the denominator under the square root goes to zero there. This requirement fixes $P^r$ to be
\begin{equation}\label{eq:Prsolutions}
    P^r = \left. \pm \, r^3e^{-\frac{w}{2}-\frac{10}{3}\chi}\sqrt{g\left(\frac{e^{2\chi+2\eta}}{9}+q^2\right)} -q\, a_4  \right|_{r = r_{min}}.
\end{equation}
To determine if we should pick the plus- or the minus-sign above, we discuss the wall profile $\xi(r)$. Following the results of \cite{Henriksson:2021zei,Bigazzi:2021ucw}, we expect this profile to have a U-shape, with the bottom of the U pointing in the direction of propagation. This follows from the fact that the derivative $\xi'(r)$ must diverge not only at $r_\text{min}$, but also at the horizon $r=r_H$, as a consequence of the overall factor of $g^{-1}$ in (\ref{eq:D5_KsiiDeriv}) ($g$ goes to zero linearly at the horizon). A necessary requirement to get a U-shape is that $\xi'(r)$ has a zero at some radius between $r_H$ and $r_\text{min}$, where the profile turns around. The only way (\ref{eq:D5_KsiiDeriv}) can have a zero turns out to be if the factor in front of the square root, $(P^r+q\, q_4)/g$, has a zero.\footnote{The numerator inside the square root will also have a zero, but this will be required to ``cancel out" another zero in the denominator of the square root, as we will explain shortly, and therefore does not lead to a zero of $\xi'(r)$.} The function $g$ is always positive and finite for $r_H<r<r_\text{min}$, so we focus on
\begin{equation}
    P^r+q\, a_4 =  \left. \pm \, r^3e^{-\frac{w}{2}-\frac{10}{3}\chi}\sqrt{g\left(\frac{e^{2\chi+2\eta}}{9}+q^2\right)}  \right|_{r = r_\text{min}} + q \left[a_4(r)-a_4(r_\text{min}) \right] \ .
\end{equation}
We note that the second term is always negative for $r_H<r<r_\text{min}$, since $a_4$ increases monotonically from zero at $r_H$, so we must pick the first, constant, term to be positive to be able to have a zero.

Having thus agreed to choose the plus sign, we can compare (\ref{eq:Prsolutions}) with  (\ref{eq:D5effPot}); if we restored the overall constant we factored out in the definition (\ref{eq:PrDefinition}), we would now have fixed $P^r$ to the value of the effective potential at its minimum.

Next, we note that the denominator in the square root of (\ref{eq:D5_KsiiDeriv}), with $P^r$ fixed in this way, always changes sign at some radius $r=r_0$ between the horizon and $r_\text{min}$. (The denominator starts out positive, equaling $|P^r|$ at the horizon, then becomes negative as it crosses zero at $r_0$, before it turns around and approaches zero \emph{from below} as $r\to r_\text{min}$.) But if we want $\xi'(r)$ to be everywhere real, this requires the numerator to also change sign exactly at $r_0$; from this we get
\begin{equation}
    \left.v^2=\frac{g\, e^{-w}}{r^2}\right|_{r = r_0}
\end{equation}
In this way, the wall speed can be fixed for each background solution and for each $q$.

\begin{figure}[H]
    \centering
    \includegraphics[scale = 0.7]{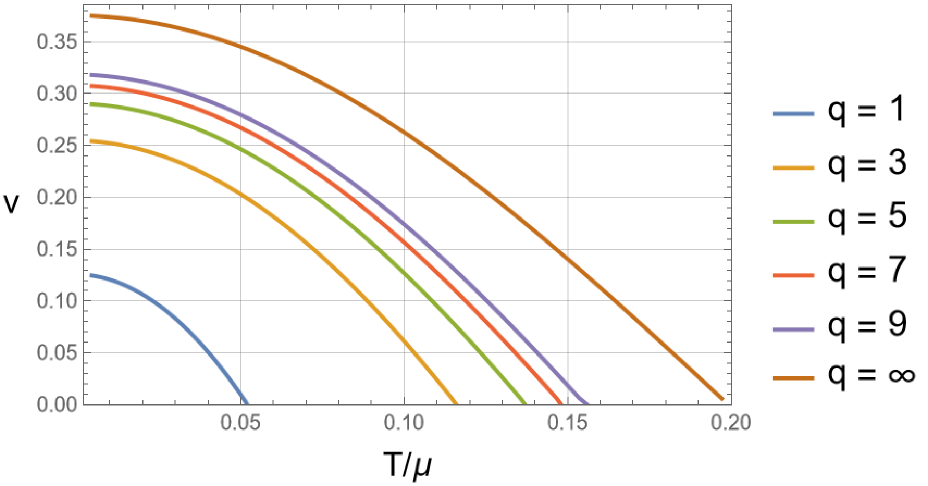}
    \caption{The terminal wall speed of the expanding bubble as a function of $T/\mu$, for different values of the flux parameter $q$. The $q=\infty$ line represents the D3-brane result from \cite{Henriksson:2021zei}.}
    \label{fig:vTmuPlot}
\end{figure}

The results are shown as a function of $T/\mu$, for a few different values of $q$, in Fig. \ref{fig:vTmuPlot}. We note that at fixed $T/\mu$, the velocity is a monotonically increasing function of $q$, asymptoting at large $q$ to the D3-brane result of \cite{Henriksson:2021zei}, as expected from the discussion around Eq. (\ref{eq:D5actionD3limit}). We note that the wall speeds are low; even in the limit of small temperatures they remain well below the sounds speed of the surrounding plasma, which takes the conformal value $v_s=1/\sqrt{3}\approx 0.577$.

\begin{figure}[H]
    \centering
    \includegraphics[scale = 0.7]{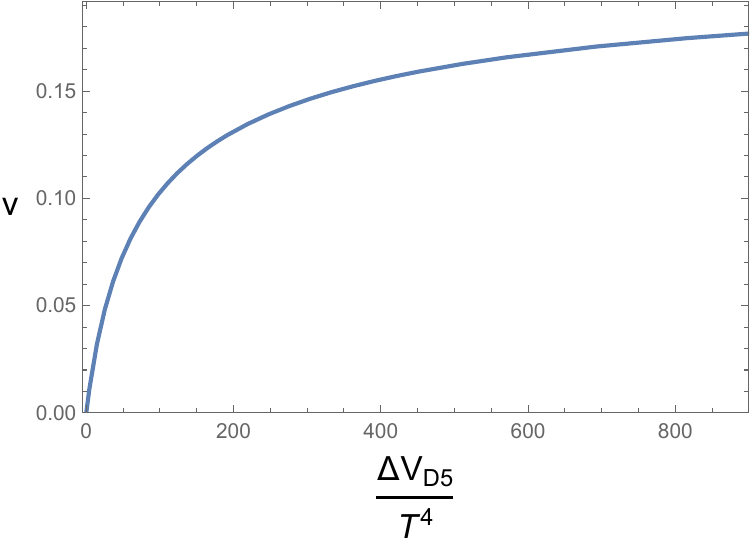}
    \caption{The wall speed as a function of the D5-brane potential energy difference between the horizon and the global minimum, at $T/\mu\approx 0.12$.}
    \label{fig:vDeltaVPlot}
\end{figure}

In earlier computations of the wall speed using holographic duality, it was noted that the terminal speed was proportional to the pressure difference between the two phases, normalized by the energy density of the phase outside the bubble \cite{Bea:2022mfb}. While later studies have argued that this is only true in some region of parameter space, it is interesting to check if something similar holds here. An easy check is to fix the background (fix $T/\mu$) and study the wall speed as a function of the flux $q$. We take the difference in the D5-brane potential (\ref{eq:D5effPot}) between the ``false vacuum" at the horizon and the ``true vacuum" at $r=r_\text{min}$ as a proxy for the pressure difference. Instead of the energy density, we normalize by $T^4$; since we have fixed the background, the normalization is either way just an overall constant. The result for $T/\mu\approx 0.12$ is seen in Fig. \ref{fig:vDeltaVPlot}, and is clearly nonlinear (similar curves result from other $T/\mu$-values). In particular, when the flux $q$ becomes very large, $\Delta V_{D5}$ continues to grow (proportionally to $q$) while the wall speed saturates to the D3-brane result. It would be interesting to see how this result would be affected by the inclusion of backreaction.

\begin{figure}[H]
    \centering
    \includegraphics[scale = 0.7]{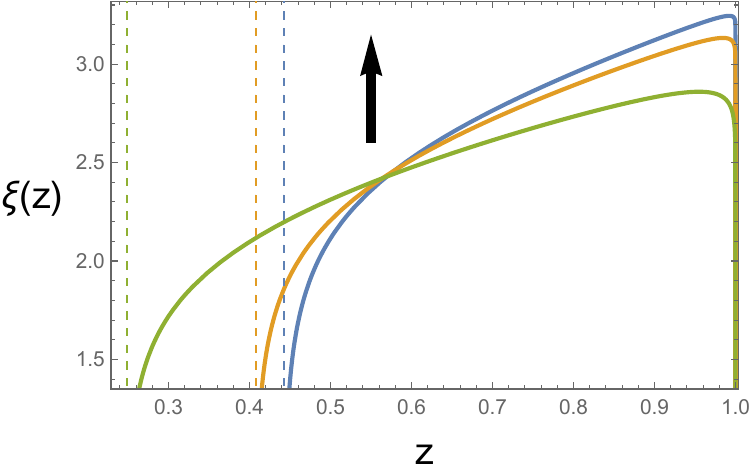}
    \caption{Example solutions for the bubble wall profile $\xi$ as a function of $z=r_H/r$. Here, $T/\mu\approx 0.12$, and the blue, orange, and green curves are branes with flux $q=4,5$, and 20, respectively. The black arrow indicates the direction of propagation of the wall. The dashed vertical lines indicates the location of the true minimum of the effective potential for the three $q$-values.}
    \label{fig:wallProfiles}
\end{figure}

Finally, having fixed $P^r$ and $v$, we can now integrate (\ref{eq:D5_KsiiDeriv}) to arrive at the wall profile $\xi(r)$. Some example solutions, for the same parameter values as the nucleating bubbles in Fig. \ref{fig:bubbleSolutions}, are shown in Fig. \ref{fig:wallProfiles}. As discussed above, we see that the solutions diverge (to minus infinity) near the horizon at $z=1$, as well as at the minima of the effective potential, which for the chosen parameters are indicated by dashed vertical lines. Note that these bubble walls should be thought of as propagating upward in the plot, as indicated by the black arrow. Also note that the intersection of the curves has been chosen at random; the curves can be shifted up and down by an arbitrary integration constant.

%%%%%%%%%%%%%%%%%%%%%%%%%%%%%%%
%%%%%%%%%%%%%%%%%%%%%%%%%%%%%%%
\section{Discussion}\label{sec:discussion}
%%%%%%%%%%%%%%%%%%%%%%%%%%%%%%%
%%%%%%%%%%%%%%%%%%%%%%%%%%%%%%%

In this paper we have studied the dynamics of bubble nucleation, and the subsequent bubble expansion, in a first-order phase transition in a gauge theory at non-zero density and temperature. The model was the superconformal Klebanov--Witten theory, in its strongly interacting regime. The calculations were done in the holographic dual, string theory on (asymptotically) AdS$_5\times T^{1,1}$. The theory transitions to a type of Higgs phase, which, in the gravity dual, is described by a D-brane localizing at a certain radius in the bulk. We focused on D5-branes with $M$ D3-branes dissolved in them, representing a breaking pattern $SU(N)\times SU(N) \to SU(N-M)\times SU(N-M-1)$.

We showed how critical bubbles, which represent fluctuations taking the system to the new phase, form as D-branes stretching out from the horizon. The action of theses bubble solutions, which we computed, provides estimates of the nucleation rate. The principal finding is that the bubble action, and thus the nucleation rate, has a minimum at a certain value for $M$ (or $q$), whose dependence on $T/\mu$ is well-fitted by Eq. (\ref{eq:qOpt}). This could be taken as a signal that the symmetry breaking pattern varies as the temperature is lowered: Near $T_c\approx 0.2\mu$, there is a preference for large $M$-values, and a fairly symmetric breaking of the two gauge groups, while at smaller temperatures the preferred $M$ decreases, leading to a larger asymmetry. However, we note that the D5-brane nucleation rates, which as discussed in section \ref{sec:nucleation} are anyway lower than the simpler D3-brane rates, need not say anything about the actual ground state, for which backreaction needs to be included.

We proceeded to study the late-time expansion of the critical bubble. The main result here is a computation of the terminal wall speed of these bubbles, which results from the equilibrium between outward pressure and energy loss through friction with the surrounding plasma. This is an exceedingly difficult quantity to compute in field theory, even at weak couplings, and is of great interest in early-universe cosmology, where it can determine whether a certain phase transition gives rise to a detectable gravitational wave signal \cite{Caprini:2019egz}. The speeds we find depends on $T/\mu$, decreasing from a maximum value at $T/\mu=0$ down to zero at the critical value $T/\mu\approx 0.2$; it also depends on the number of dissolved D3-branes $M$, increasing from zero at the critical $M$-value needed for the instability, to the D3-brane results from \cite{Henriksson:2021zei} as $M\to\infty$. The speed always remains below 0.4, well below the value of the conformal sound speed $1/\sqrt{3}$. The process could thus be classified as a deflagration, similarly to other results for thermal phase transitions from holographic duality \cite{Bea:2021zsu,Henriksson:2021zei}.\footnote{If a zero-density FOPT extends down to zero temperature, the wall speed always approaches 1, also in holographic models \cite{Bigazzi:2021ucw}.} It would be interesting to extend our analysis to holographic backgrounds which will have stiff equations of state giving rise to speeds of sound exceeding the conformal value~\cite{Hoyos:2016cob,Ecker:2017fyh}.

We now note some interesting directions for future research. To begin with, the current paper has only looked at the moment of nucleation and the steady-state expansion at asymptotic times. It would be interesting to connect them together by simulating the expansion of the bubble in real-time. This would require the numerical solution of the partial differential equation in time and radial coordinate, similar to what has recently been done in \cite{Evans:2024ilx}.

Further in the context of real-time simulations, one could try to capture the nucleation event itself. This has been done in standard field theory systems, both using Hamiltonian deterministic evolution as well as stochastic Langevin dynamics; see for example \cite{Pirvu:2024nbe} and references therein. The results of \cite{Pirvu:2024nbe}, in particular, are interesting since they find a discrepancy with the standard Euclidean calculation. They attribute this to slow relaxation at weak coupling, and it would therefore be worthwhile to compare this with a strongly coupled system such as in the present paper. In our case, one could imagine placing the probe brane in a thermally excited state near the black brane horizon and studying its time evolution. Since the gravitational system is highly dissipative --- the brane would fall towards the horizon --- one must likely add stochastic source terms, giving a Langevin-type equation, in order to see the nucleation. These stochastic terms would originate from quantum effects, for example the modes on the brane being excited by the thermal environment of the horizon \cite{deBoer:2008gu}. One could attempt to compute these from first principles, or add them in a phenomenological manner while trying to satisfy the fluctuation-dissipation theorem.

One simple extension would be to redo these computations of nucleation rates and wall speeds in more general systems; for example, the KW gauge theory admits various deformations, one of which was studied in \cite{Henriksson:2019ifu}. One could also look for similar instabilities in other dimensions, such as the $(2+1)$-dimensional field theory in \cite{Faedo:2017fbv,Faedo:2023nuc}.

Finally, we have the perhaps most pressing question: What is the end state of this instability, assuming it exists? While the D5-brane potential has a minimum at a finite radial position, the D3-brane potential has its minimum at infinite radius, at the AdS boundary. There is thus a risk that no completely stable end state exists; all states we can construct might be unstable to further D3-brane nucleation, until all branes have moved infinitely far away. This cannot be completely resolved without adding backreaction. This has been partly accomplished in a bottom-up model (designed to mimic features of the top-down case) in \cite{Henriksson:2022mgq}, and in case of $\mathcal{N}=4$ super-Yang-Mills theory in \cite{Kim:2023sig,Choi:2024xnv}. In those cases, the stable ground state only exist when the theory is placed on a three-sphere --- this lifts the moduli space, resulting in the brane effective potential having a minimum at finite radius. Since in the conifold case the D5-branes have their minimum at finite radius, one might hope that a backreacted solution involving D5-branes could completely escape the brane nucleation instability. The D5-branes would sit at some finite radius (or radii), and could perhaps be smeared over the remaining internal directions such that they form a co-dimension one shell. Such a solution, if it exists, would extrapolate between the KW theory in the UV and some variation of the Klebanov--Strassler theory (at non-zero baryon density) in the IR. We leave these interesting problems for the future.

\paragraph{Acknowledgements.}%---%
We acknowledge useful discussions with Carlos Hoyos. O. H. has been supported in part by the Waldemar von Frenckell foundation. N.~J. has been supported in part by the Research Council of Finland grant no. 354533.

%%%%%%%%%%%%%%%%%%%%%%%%%%%%%%%
%%%%%%%%%%%%%%%%%%%%%%%%%%%%%%%

\bibliographystyle{apsrev4-1}
\bibliography{bibliography}

\end{document}